\documentstyle[preprint,aps]{revtex}
\tighten
\input psfig.tex
\begin{document}
\draft

\title{Correlation decay and conformal anomaly in the two-dimensional random-bond Ising ferromagnet}
\author{ S. L. A. de Queiroz\cite{email} }
\address{
Instituto de F\'\i sica, Universidade Federal Fluminense,\\ Outeiro de S\~ao Jo\~ao Batista s/n, 24210-130 Niter\'oi RJ, Brazil}
\date{\today}
\maketitle
\begin{abstract}
 The two-dimensional random-bond Ising model is numerically studied on long strips by transfer-matrix methods. It is shown that the rate of decay of correlations at criticality, as derived from averages of the two largest Lyapunov exponents
plus conformal invariance arguments, differs from that obtained through direct evaluation of correlation functions. The latter is found to be, within error bars, the same as in pure systems. Our results confirm field-theoretical predictions.
The conformal anomaly $c$ is calculated from the leading finite-width correction  to the averaged free energy on strips. Estimates thus obtained are consistent with $c=1/2$, the same as for the pure Ising model.
 
\end{abstract}

\narrowtext
\newpage

Conformal invariance methods have been important in the study of critical
properties of two-dimensional spin models\cite{cardy}, providing a number of
exact results. Most applications have been to pure (translationally invariant) systems. When  disorder is introduced, invariance under conformal transformations at criticality can be said to be preserved only in an average
sense. This raises the question of whether averaged properties still connect to  quantities in conformal theory, in the same way as their counterparts
for pure systems. Further, it is well known that performing averages in inhomogeneous systems is a subtle matter\cite{derrida,crisanti,ranmat}, owing to the fact that the moments of local quantities scale with independent exponents.
This is in contrast with the non-random case, where one has constant gap indices.

In the present work, we study properties of the two-dimensional random-bond
Ising model with a binary distribution of ferromagnetic interaction strengths,
each occurring with equal probability. As the transition temperature is exactly
known from duality\cite{fisch,kinzel}, one can be sure that numerical errors due to imprecise knowledge of the critical point are absent.
In strip calculations, the only sources of such errors will then be the finite
strip width and, for a random system, those arising from the averaging process.
The former can be dealt with by finite-size scaling theory\cite{fs1,fs2}, whereas one expects
to reduce the effects of the latter by studying large enough samples. Averages
of the following quantities are calculated at criticality: $(i)$ the two largest Lyapunov exponents; $(ii)$ spin-spin correlation functions; and $(iii)$ free energy per spin. From $(i)$ and conformal invariance arguments, we obtain what
is supposed to represent the $typical$ (as opposed to $average$)\cite{derrida,crisanti,ranmat,ludwig} decay of the correlation functions. As shown below, results from $(i)$ and $(ii)$ differ in a way that
corroborates the field-theoretic arguments of Ref. \onlinecite{ludwig}. Numerical data from $(iii)$ are used to estimate the leading finite-width correction to the bulk free energy, which is known\cite{bcn,aff} to be proportional to the conformal anomaly.

We have used long strips of a square lattice, of width $2 \leq L \leq 13$ sites
with periodic boundary conditions. In order to provide samples that are
sufficiently representative of disorder, we iterated the transfer matrix\cite{fs2} typically along $10^5$ lattice spacings. At each step, the respective vertical and horizontal bonds between first-neighbour spins $i$ and $j$ were drawn from a probability distribution 
\begin{equation}
 P(J_{ij})= {1 \over 2} ( \delta (J_{ij} -J_0) +  \delta (J_{ij} -rJ_0) ) \ \ \ ,\ \ 0 \leq r \leq 1 \ \ ,
\label{eq:1}
\end{equation}
\noindent which ensures\cite{fisch,kinzel} that the critical temperature $\beta_c = 1/k_B T_c$ \ of the corresponding
two-dimensional system is given by
\begin{equation}
\sinh (2\beta_{c} J_{0})\sinh (2\beta_{c}r J_{0}) = 1 \ \ .
\label{eq:2}
\end{equation}
Two values of $r$ were taken in calculations: $r=0.5$ and $0.01$. These roughly 
represent `weak' and `strong' disorder, respectively. The critical temperatures,
from Eq. \ref{eq:2}, are: $T_c\, (0.5)/J_0 = 1.641 \dots$; $T_c\, (0.01)/J_0 = 0.5089 \dots$ (to be compared with  $T_c\, (1)/J_0 = 2.269 \dots$). The choice of $r=0.5$ allowed us to compare our results with
those of Glaus\cite{glaus}, who obtained averages of the two largest Lyapunov exponents and of the free energy for strips of $L \leq 8$. The reason for fixing $r=0.01$ was that, in site-diluted Ising systems\cite{dqrbs}, a transfer-matrix $ansatz$ shows that for concentrations such that $T_c \simeq 0.2 T_c(1)$ one
already is crossing over towards percolative behaviour. Analogy thus suggested  
that, for a critical temperature of the same order in our case, one might be
able to detect similar effects (if they are present at all). 
\bigskip

The procedure for evaluation of the largest Lyapunov exponent $\Lambda_{L}^{0}$ for a strip of width $L$ and length  $N \gg 1$ is well known\cite{ranmat,glaus}.
Starting from an arbitrary
initial vector ${\bf v}_0$, one generates the transfer matrices ${\cal T}_{i}$
that connect columns $i$ and $i+1$, drawing bonds from the distribution 
Eq. \ref{eq:1}, and applies them successively, to obtain:
\begin{equation}
 \Lambda_{L}^{0} = {1 \over N} \ln \Biggl\{ {\Bigl\Vert \prod_{i=1}^N {\cal T}_{i} {\bf v}_0 \Bigr\Vert \over \bigl\Vert {\bf v}_0 \bigr\Vert }\Biggr\}\ \ .
\label{eq:3}
\end{equation}
The average free energy per site is, in units of $k_{B}T$:
\begin{equation}
 f_{L}^{\ ave}(T) = - {1 \over L} \Lambda_{L}^{0}\ \ .
\label{eq:4}
\end{equation}
The second, third $etc.$ exponents can generally be obtained by iteration of
a set of initial vectors, ${\bf v}_i$, orthogonal to each other and to 
${\bf v}_0$; care must be taken to orthogonalise the iterated vectors every few
steps\cite{ranmat}, to prevent contamination with dominant-exponent components. Here, the task is easier since we shall be concerned only with the two largest Lyapunov exponents, and  the Hamiltonian is invariant
under global spin inversion. In order to calculate $\Lambda_{L}^{0}$
($\Lambda_{L}^{1}$), it is then sufficient to iterate ${\bf v}_0$ (${\bf v}_1$)
which is even (odd) under that same symmetry\cite{fs2,glaus}, with no need for intermediate decontamination of the iterates of ${\bf v}_1$.

From $\Lambda_{L}^{0}$ and $\Lambda_{L}^{1}$ one can form estimates of
\begin{equation}
\xi_{L}^{-1} =  \Lambda_{L}^{0} - \Lambda_{L}^{1} \ \ ,
\label{eq:5}
\end{equation}
\noindent which, at the critical point, is related to the typical decay of correlations on a fixed sample in the plane\cite{ludwig}. The correlation functions in this case are predicted\cite{ludwig} to have the form
\begin{equation}
 < \sigma_0 \sigma_R > \sim R^{-1/4} (\Delta \ln R)^{-1/8} \ \ \ {\rm (fixed\  sample)},
\label{eq:6}
\end{equation}
\noindent for $\ln (\Delta \ln R)$ large, where $\Delta$ is proportional to the intensity of disorder. If now one tries to fit the standard relation\cite{cardy}
between correlation length at criticality and the correlation-decay exponent $\eta$:
\begin{equation}
\eta =L/\pi \xi_{L} \ \ ,
\label{eq:7}
\end{equation}
\noindent which holds for pure systems, Eq. \ref{eq:6} implies that the result
will in fact be an effective exponent, larger than $1/4$ by an amount which increases with increasing disorder. As shown below, this is what arises from our numerical study.   

The direct calculation of correlation functions follows the lines of Section 1.4
of Ref. \onlinecite{fs2}, with standard adaptations for an inhomogeneous system. For two spins in, say, row 1, separated by a distance $R$, and for a given configuration $C$ of bonds, one has:  
\begin{equation}
 < \sigma_{0}^1 \sigma_{R}^1 >_C =  \sum_{\sigma_{0} \sigma_{R}}\tilde\psi (\sigma_{0})\, \sigma_{0}^1\ \bigg(\prod_{i=0}^{R-1} {\cal T}_{i}\bigg)_ {\sigma_{0} \sigma_{R}}\ \sigma_{R}^1\, \psi (\sigma_{R}) \bigg/  \sum_{\sigma_{0} \sigma_{R}}  \tilde\psi (\sigma_{0})\ \bigg( \prod_{i=0}^{R-1} {\cal T}_{i}\bigg)_{\sigma_{0} \sigma_{R}}\ \psi (\sigma_{R})\ \ ,
\label{eq:8}
\end{equation}
\noindent where $\sigma_{0} \equiv \{ \sigma_{0}^1 \ldots \sigma_{0}^L \}$ and
correspondingly for $\sigma_{R}$; the bonds that make the transfer matrices ${\cal T}_{i}$ belong to $C$.  For pure systems the $2^L-$component vectors  $\tilde\psi$, $\psi$ are determined by the boundary conditions; for example, the choice of dominant left and right eigenvectors gives the correlation function in an infinite system\cite{fs2}. Here, one needs only be concerned with avoiding start-up effects, since there is no convergence of
iterated vectors. This is done by discarding the first few hundred
iterates of the initial vector ${\bf v}_ 0$. 
From then on, one can shift the dummy origin of Eq. \ref{eq:8} along the strip,
taking $\tilde\psi$ ($\psi$) to be the
left-- (right--) iterate of ${\bf v}^{T}_ 0$ (${\bf v}_ 0$~) up to the shifted
origin.  

One performs averages of the correlation
function (to be denoted by $\overline{< \sigma_{0}^1 \sigma_{R}^1 >}$\ ) by
shifting the origin along the strip and accumulating the corresponding results for Eq. \ref{eq:8}. Such averages behave as 
\begin{equation}
\overline{< \sigma_{0}^1 \sigma_{R}^1 >} \sim \exp (-R/\xi^{\ ave}) \ \ .
\label{eq:9}
\end{equation}
For pure systems, $\xi^{\ ave}$ coincides with the definition of $\xi_{L}$ in
Eq. \ref{eq:5}. Here we shall show that, at $T_c$ one gets (within reasonable error bars) 
\begin{equation}
\xi^{\ ave}=L/\pi \eta  \ \ ,
\label{eq:10}
\end{equation}
\noindent where $\eta=1/4$ is the same as for the pure system, with no logarithmic corrections such as those for typical correlations
on a fixed sample (Eq. \ref{eq:6}). This is again in agreement with the predictions of Ref. \onlinecite{ludwig}.
\medskip

The total strip length used
was $N = 10^5$ for $L = 2-11$ and $5 \times 10^4$ for $L=12$ and $13$ (except for free energy calculations, where $N = 10^5$ was used also for the largest widths). Evaluation, both of Lyapunov exponents and of averaged correlation functions, involves iterating initial arbitrary vectors.
In order to get rid of start-up 
effects, the first $N_0 = 2,000$ iterations were discarded. For Lyapunov exponents, the accumulated averages were recorded every
200 subsequent steps\cite{glaus}. From this set we extracted estimates of exponents
and corresponding fluctuations. This procedure was repeated three times
with different initialisations. Final values of exponents and error bars were taken respectively as arithmetic averages of estimates at the end of each run (that is, each taking into account $N-N_0$ steps), and root-mean-square averages of deviations (estimated from the sampling described above). We calculated correlation functions for distances $R$ in Eq. \ref{eq:8} ranging from 10 to 200 lattice spacings. The shifted origins were taken 200 columns apart, thus for each of the three independent runs, and each distance, one would have close to 500 samples of correlation functions (250 for
$L=12$ and $13$). As all correlations are ferromagnetic, their fluctuations are 
well-behaved (compared $e.g.$ to random-field problems, where frustration effects induce large-amplitude deviations from averages\cite{rbsmdq} ). 

Our results are depicted in Figs. \ref{fig:eta} $(a)$ and $(b)$,
respectively for $r = 0.5$ and $0.01$. We have plotted values
of $L/\pi \xi$, both with $\xi=\xi_{L}$ given by Eq. \ref{eq:5} and 
$\xi=\xi^{\ ave}$
of Eq. \ref{eq:9}, against $1/L^2$. The choice of horizontal axis is inspired by the pure case\cite{dds}, where it can be shown analytically that the leading finite-size corrections to $\eta$ are proportional to $1/L^2$.
\medskip
 
For weak disorder
$r = 0.5$,  Fig. \ref{fig:eta} $(a)$, both sequences are linear in $1/L^2$ to a good extent. Data from averaged correlation functions point inequivocally
towards the pure-system value $\eta = 1/4$, in accordance with the predictions of Ref. \onlinecite{ludwig}, and also with very recent large-scale Monte Carlo
simulations\cite{talapov}. Owing to the non-negligible fluctuations for large $L$, it is not feasible to try extrapolations of data for that region alone, as is done when no uncertainties are present\cite{dst}. Instead, we turn to global fitting procedures.  The analysis of least-squares fits of data (taking
error bars into account) from
$L = L_0$ to $L=13$, with $L_0 = 2, \ldots 10$ indicates that $\chi^2$ stabilises at $\sim 0.19$ for $L_0 = 4-6$; estimates of the extrapolated exponent and of errors are also very similar in this range. For these values of  $L_0$ a rough balance is attained, between the trend towards reducing errors by including fewer data, and that towards increasing uncertainties by giving more relative weight to 
fluctuations in individual data points. The exponent estimate from such fits is
$\eta = 0.250 \pm 0.002$. 

Turning to the sequence of Lyapunov exponents for $r = 0.5$, an analysis of least-squares fits
similar to that just described gives $\eta = 0.257 \pm 0.002$, with an even
smaller $\chi^2 \sim 0.05$. If, following Ref. \onlinecite{glaus}, one
tries to plot data against $1/L$, the amount of curvature is found to increase substantially. Taken on their own, the present results would
seem to characterise a non-universal exponent, with leading finite-size
corrections depending on $1/L^2$.
When data for strong disorder are considered, however, this picture is not confirmed. As we shall see below, an overall scenario emerges in which
the above estimate is to be regarded as an effective exponent, signalling the presence of logarithmic corrections of Eq. \ref{eq:6}.  
\medskip
 
For strong disorder
$r = 0.01$,  Fig. \ref{fig:eta} $(b)$, the sequence of data from averaged correlation functions points again close to $\eta = 1/4$, though with a larger
amount of scatter than in the previous case. Consideration of least-squares fits, analogously to that done above, leads to 
$\eta = 0.235 \pm 0.015$. The corresponding $\chi^2$ is $\sim 1.8$.

Estimates of $\eta$ from Lyapunov exponents for $r = 0.01$ obviously do not scale linearly with $1/L^2$; one finds that the best linear fit is against $L^{-\phi}$ with $\phi \simeq 0.2$,
in which case the extrapolate $\eta (L \rightarrow \infty)$ is $\sim 0$. We
refrain from attaching much significance to the latter result, as extrapolations
with such low powers of $1/L$ are rather unreliable. Instead, we recall
that, for a given finite $L$,  estimates of $\eta$  are always larger than those for weak disorder. This is consistent with the logarithmic corrections for fixed-sample correlations displayed $e.g.$ in Eq. \ref{eq:6}. Analytical expressions for the correlation
length of energy-energy correlation functions\cite{cardy2,lc}, though not
directly comparable with the present results, also show corrections in inverse powers of $\ln L$ to the pure-system behaviour, whose absolute value increases with disorder.   
\medskip

We now present results for the conformal anomaly. Table 1 shows the negative free energy
per site, in units of $k_BT$, as obtained from Eq. \ref{eq:4}, at the respective critical temperatures given by Eq. \ref{eq:2} for $r=1$ (pure system), $0.5$ and
$0.01$. Data for $r=1$ are from Bl\"ote and Nightingale\cite{bn}, and shall be useful for comparison among different fitting procedures. The conformal anomaly
$c$ is related to the finite-width free energy at criticality by\cite{bcn,aff}:
\begin{equation}
f_{L}(T_c) =  f_{\infty}(T_c)-{\pi c \over 6L^2 } + \dots\ \ .
\label{eq:11}
\end{equation}
We begin by recalling
that, for pure Ising systems, finite-size estimates of $c$ may approach the exact value $c=1/2$ either from above or below, depending on what
sort of corrections to the free energy are incorporated. If one stops at order
$1/L^2$, the sequence of two-strip approximations with $L$ and $L+1$ reaches $1/2$ from above; plugging in the next term (shown numerically by  Bl\"ote and Nightingale\cite{bn} to be $\propto 1/L^4$), convergence of three-strip ($L$,~
$L+1$,~$L+2$) estimates is from below and much faster. This is illustrated in columns $(i)$ and $(ii)$ of Table 2.  The same happens when one performs straight-line
or parabolic least-squares fits of data against $1/L^2$, using data from 
$L = L_{min} = 2, \dots 11$ up to $L_{max} = 13$, as shown in columns $(iii)$ and $(iv)$ of that table.  

When disorder is considered, Table 1 shows that
the error bars are a serious obstacle to evaluation of estimates from pairs or triplets of strip widths, especially for large $L$. In order to produce reliable results, one would need free-energy errors typically smaller than one part in $10^5$ for $L \gtrsim 10$. The strips used would have to be respectively $\sim 10^2$ (for $r=0.5$) and $10^4$ (for $r=0.01$) times longer than the present value $N=10^5$, assuming fluctuations to be normally distributed. Though not unfeasible, such task would demand a considerable amount of computer time.
We shall then try to extract information from comparison between least-squares fits of disordered-system data and their pure-system counterparts. In Table 3
are shown results of linear and parabolic fits against $1/L^2$, both for $r=0.5$ and $0.01$. In general, we found that fits starting at values of $L$ larger than
those displayed gave such large errors as to be rendered meaningless. For weak disorder, columns $(i)$ and $(ii)$, the overall trends closely resemble those found in the pure case of Table 2, including the approach of central estimates to $c=1/2$ (from above or below, depending only on the type of fit),
though of course with much larger errors.
Results in column $(i)$ are essentially the same as in  Ref. \onlinecite{glaus}.
As expected, error bars are even larger for strong disorder, columns $(iii)$ and $(iv)$; though straight-line fits deteriorate very quickly, one still has 
reasonably well-behaved estimates from parabolas. However, if monotonic approach to $c=1/2$ from below is present in the latter sequence, it is utterly smeared out by  fluctuations. 
\bigskip

Our results can be summarised as follows.
\smallskip
\par\noindent $(i)$ Averaged correlation functions at criticality of the disordered system decay with pure power-law behaviour,
the same exponent $\eta = 1/4$ of the pure
Ising model. This is contrary to early results, according to
which disorder would lead to $\eta = 0$\cite{dotsenko}, but
supports field-theoretical\cite{ludwig} and Monte-Carlo\cite{talapov} evidence. Similarly, very recent results for site-diluted systems\cite{kim,kuhn} indicate that, although susceptibility ($\gamma$) and correlation length ($\nu$) exponents seem to be concentration-dependent, the ratio $\gamma/\nu = 2 - \eta$  remains at $7/4$ .
\smallskip
\par\noindent $(ii)$ Typical (as opposed to averaged) correlation decay,
as obtained from the two largest Lyapunov exponents, behaves consistently with the presence of (non self-averaging) logarithmic corrections\cite{ludwig}.        \smallskip
\par\noindent $(iii)$ Conformal anomaly estimates for the disordered Ising model behave in the same way (within error bars) as the corresponding ones for
pure systems. We are thus led to state that no evidence has been found that
$c \not= 1/2$. However, our results are not accurate enough to detect logarithmic corrections predicted by field theory\cite{cardy2}.
\medskip

One might view the estimate $\eta = 0.235 \pm 0.015$, from averaged  correlation
functions for $r=0.01$, as signalling a crossover towards percolation-like
behaviour (for which\cite{dds,denn} $\eta_p = 5/24$). Though this is indeed
expected\cite{dqrbs,rbs} to occur sufficiently close to $T=0$ ($i.e.$, $r=0$),   a systematic investigation of several points in that region is
needed in order to sort out finite-size effects from $r$--dependent ones.   
We intend to do so, as part of a forthcoming, broader-ranging study\cite{unpub} which will include generalisations of a calculational procedure initially devised for site-diluted problems\cite{dqrbs} to random-bond and random-field systems . 
 
\acknowledgements 
The author thanks the following persons and institutions: F. C. Alcaraz, J. L.
Cardy and R. B. Stinchcombe for useful discussions; Department of Theoretical Physics, Oxford, for hospitality;
the cooperation agreement between Academia Brasileira de Ci\^encias and
the Royal Society for funding his visit to Oxford in early 1994; Departamento de F\'\i sica, PUC/RJ for use of their computational facilities, and
Brazilian agencies  Conselho Nacional 
de Desenvolvimento Cient\'\i fico e Tecnol\'ogico and Financiadora de Estudos e Projetos, for financial support.


\newpage
\vskip 4.0cm
\begin{table}
\caption{
Critical free energies per site. Uncertainties in last quoted digits are shown in parentheses.}
\vskip 0.7cm 
 \halign to \hsize{\hskip 4.0cm\hfil#\quad\hfil&\quad\hfil#\quad\hfil&\hfil#\quad\hfil&\hfil#\quad\hfil\cr
    $L$ & $r=1$ &  0.5 &  0.01 \cr \noalign{\smallskip}
  2  & 1.886426125762 & 1.0390 (6) & 2.121 (4) \cr
  3  & 1.842546256346 & 0.9956 (6) & 2.081 (4) \cr
  4  & 1.828157728044 & 0.9813 (5) & 2.068 (4) \cr
  5  & 1.821819028739 & 0.9750 (4) & 2.062 (3) \cr
  6  & 1.818468940405 & 0.9716 (6) & 2.059 (4) \cr
  7  & 1.816478784614 & 0.9695 (7) & 2.056 (5) \cr
  8  & 1.815198160430 & 0.9683 (4) & 2.055 (3) \cr
  9  & 1.814324882545 & 0.9675 (4) & 2.055 (3) \cr
  10 & 1.813702482337 & 0.9669 (5) & 2.054 (4) \cr
  11 & 1.813243149887 & 0.9663 (4) & 2.053 (3) \cr
  12 & 1.812894445034 & 0.9661 (4) & 2.053 (3) \cr
  13 & 1.812623456975 & 0.9657 (3) & 2.052 (3) \cr
}
\end{table}

\vskip 2.0cm
\begin{table}
\caption{
Conformal anomaly estimates for pure Ising model.
 $(i)$ : two-point fits ($L,L+1$) with $1/L^2$ corrections.
 $(ii)$ : three-point fits ($L,L+1,L+2$) with $1/L^2$ and $1/L^4$ corrections.
 $(iii)$ : straight-line least-squares fits against $1/L^2$  with data from $L$ to 13.
 $(iv)$ : parabolic least-squares fits against $1/L^2$  with data from $L$ to 13.
Uncertainties in last quoted digits are shown in parentheses.}
\vskip 0.7cm 
 \halign to \hsize{\hskip 3.0cm#\quad\hfil&\quad#\quad\hfil&#\quad\hfil&#\quad\hfil&#\quad\hfil\cr
    $L$ & $(i)$ &  $(ii)$ &  $(iii)$ & $(iv)$  \cr \noalign{\smallskip}
  2 &   0.60339151 &  0.53003814 & 0.5750 (55) & 0.51113 (211)   \cr
  3 &   0.56530418 &  0.49875466 & 0.5408 (39) & 0.49741 (13)  \cr
  4 &   0.53804550 &  0.49507433 & 0.5230 (23) & 0.49722 (27)  \cr
  5 &   0.52348887 &  0.49673455 & 0.5146 (14) & 0.49836 (18)  \cr
  6 &   0.51575527 &  0.49823978 & 0.5104 (9)  & 0.49909 (10)  \cr
  7 &   0.51133776 &  0.49906829 & 0.5080 (6)  & 0.49947 (5)  \cr
  8 &   0.50859245 &  0.49948159 & 0.5065 (4)  & 0.49966 (3)  \cr
  9 &   0.50676022 &  0.49969077 & 0.5055 (3)  & 0.49977 (2)  \cr
  10 &  0.50546906 &  0.49980307 & 0.5048 (2)  & 0.49983 (1)  \cr
  11 &  0.50452116 &  0.49986769 & 0.5042 (2)  & 0.49986769   \cr
  12 &  0.50380296 &             & 0.50380296  &              \cr
}
\end{table}

\vskip 2.0cm
\begin{table}
\caption{
Conformal anomaly estimates for bond-disordered Ising model.
 $(i)$ :  straight-line least-squares fits against $1/L^2$  with data from $L$ to 13; $r=0.5$.
 $(ii)$ : parabolic least-squares fits against $1/L^2$  with data from $L$ to 13;  $r=0.5$. 
 $(iii)$ : straight-line least-squares fits against $1/L^2$  with data from $L$ to 13; $r=0.01$.
 $(iv)$ : parabolic least-squares fits against $1/L^2$  with data from $L$ to 13; $r=0.01$.
Uncertainties in last quoted digits are shown in parentheses.}
\vskip 0.7cm 
 \halign to \hsize{\hskip 3.0cm#\quad\hfil&\quad#\quad\hfil&#\quad\hfil&#\quad\hfil&#\quad\hfil\cr
    $L$ & $(i)$ &  $(ii)$ &  $(iii)$ & $(iv)$  \cr \noalign{\smallskip}
  2 &   0.569 (4)  &  0.511 (3)  & 0.53 (3)  & 0.506 (10)  \cr
  3 &   0.536 (10) &  0.498 (3)  & 0.52 (7)  & 0.508 (20)  \cr
  4 &   0.521 (16) &  0.499 (6)  & 0.51 (12) & 0.533 (45)  \cr
  5 &   0.516 (23) &  0.495 (13) & 0.52 (17) & 0.503 (90)  \cr
  6 &   0.509 (46) &  0.492 (26) & 0.51 (34) &          \cr
  7 &   0.502 (68) &          &        &          \cr
}
\end{table}

\begin{figure}
\caption{
$\eta \equiv L/\pi \xi$, with  $\xi=\xi_{L}$ of Eq. 5 (triangles) and $\xi=\xi^{\ ave}$ of Eq. 9 (circles) against $1/L^2$. The square on the vertical axis is at the pure system value $\eta=1/4$ . Straight lines are least-square fits for $L= 4-13$. $(a)$ : $r=0.5$; $(b)$ : $r=0.01$ . }
\label{fig:eta}
\end{figure}

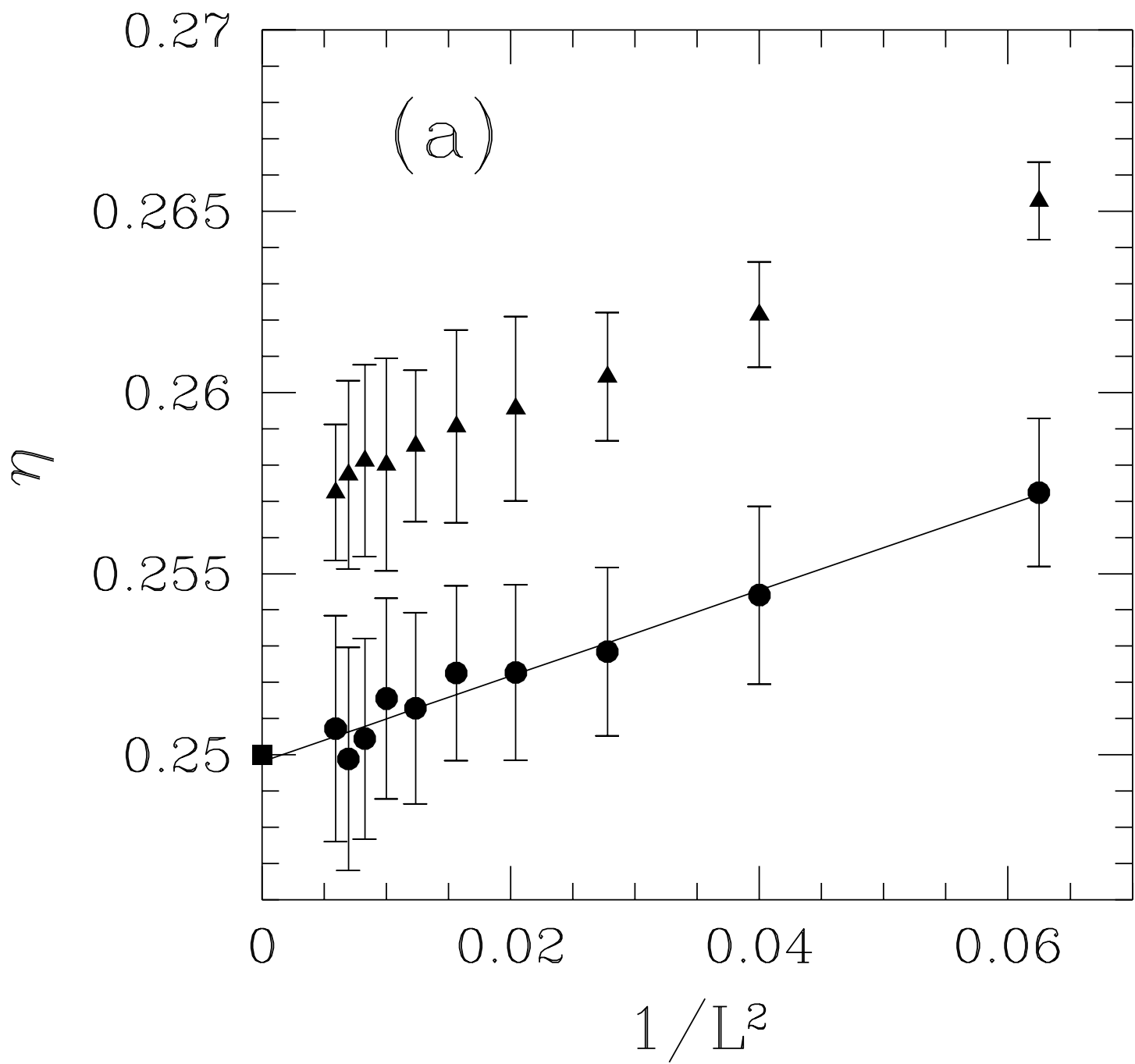
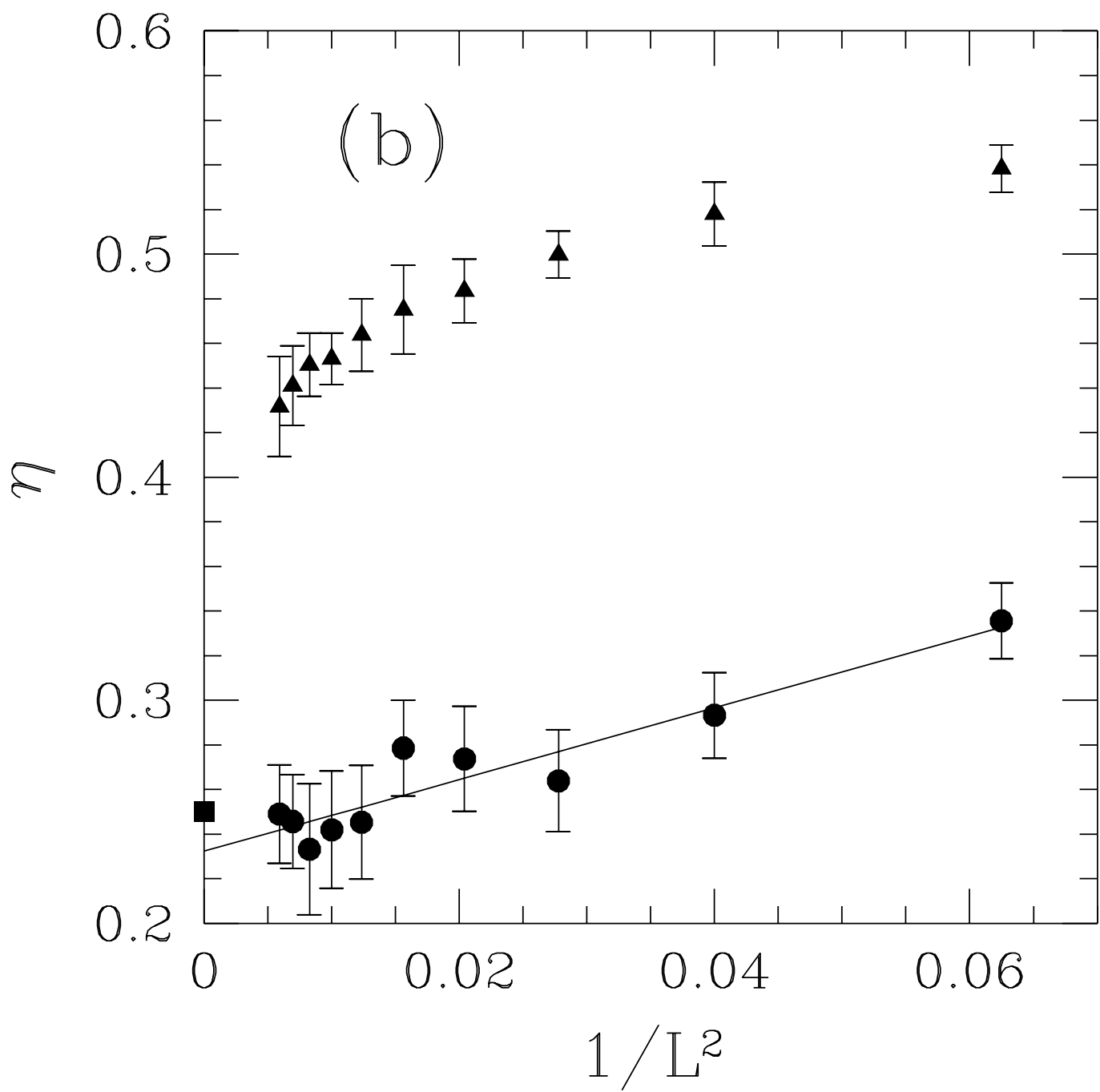
\end{document}